\documentclass[final,5p,times,twocolumn,sort&compress]{elsarticle}
\usepackage{booktabs}
\usepackage{amsmath}
\usepackage{enumitem}
\setlist{nosep}
\usepackage[colorlinks, citecolor=blue, urlcolor=blue, linkcolor=blue, bookmarksdepth=3]{hyperref}
\usepackage{graphicx}
\graphicspath{{./fig/}}
\usepackage{listings}
\journal{Computer Physics Communications}

\begin{document}

\begin{frontmatter}

\title{GEARS - A Fully Run-Time Configurable Geant4 Application}
\author{Jing Liu}
\ead{jing.liu@usd.edu}
\affiliation{organization={Department of Physics, University of South Dakota},
            addressline={414 E. Clark St.}, 
            city={Vermillion},
            postcode={57069},
            state={SD},
            country={USA}}

\begin{abstract}
The Geant4 toolkit is the standard for simulating the passage of particles through matter, but its conventional architecture often requires users to modify and recompile C++ code to alter fundamental simulation parameters such as geometry, physics list, and primary particle source. This architectural constraint introduces significant friction for new users and slows down the experimental iteration cycle. This paper introduces GEARS (Geant4 Example Application with Rich features yet Small footprint), a universally applicable Geant4 application that fundamentally addresses this issue. GEARS achieves complete simulation configurability without C++ recompilation by strictly utilizing external configuration methods: Geometry is defined via simple text-based configuration, the Physics List is selected via the standard \verb|PHYSLIST| environment variable, and the Primary Source is defined through the General Particle Source (GPS) macro commands. Furthermore, regarding GEARS as an application instead of a framework, key features include a flat Ntuple structure with short variable names for highly efficient analysis via \verb|TTree::Draw()| and a solution for capturing vital initial step data. Output creation is also fully managed via run-time macro commands and volume properties. The project is distributed as a ready-to-use Docker container to eliminate compilation barriers. Through these design considerations, GEARS transforms Geant4 into a practical, ready-to-use tool, enabling users to rapidly prototype and execute simulations for diverse experiments solely through simple text configuration files, without ever needing to modify or compile the underlying C++ source code.
\end{abstract}


\end{frontmatter}

\section{Introduction}
Geant4 (GEometry ANd Tracking)~\cite{g403,g406,g416} is a comprehensive software toolkit written in C++ for the Monte Carlo (MC) simulation of particle transportation in matter. It is an indispensable tool in modern experimental physics, astrophysics, medical physics, and engineering, providing detailed physics models and a flexible architecture for modeling complex detector geometries.

Accessing the power of the Geant4 toolkit requires developing a custom Geant4 application. Traditionally, this introduces significant software overhead. For instance, the simplest example shipped with the toolkit, examples/basic/B1/, already involves 13 C++ source files, two compiling systems (CMake and GNU Make), and about 500 lines of code. Crucially, this basic example demonstrates only a very small fraction of the functionalities available in Geant4. This intrinsic complexity means that the learning curve is often too steep for an average beginner, as the time investment required to learn the extensive Geant4 component architecture and its required C++ abstract classes is a major obstacle, leading to rapid prototyping being severely hindered by the accompanying software engineering burden and the need for frequent recompilation.

To address this challenge and simplify the simulation workflow, GEARS~\cite{gears} has been developed. It is an application designed to maximize the utility of existing Geant4 components, making it a highly effective and production-ready platform for MC simulation, allowing users to focus their effort on experiment design and physics analysis instead of software engineering.

In stark contrast to conventional applications, the core of GEARS is contained within a single C++ source file with 555 lines of C++ code. Crucially, approximately 300 lines of this code are dedicated solely to an advanced optical properties extension (detailed in Section~\ref{s:det}). This means the core universal application, stripped of this advanced feature, is closer to 300 SLOC. The compilation process is reduced to just a few second on a regular PC. This remarkably minimal C++ footprint is the result of a deliberate computational design choice: offloading all experiment-specific definitions to run-time configuration files and macro commands. Critically, GEARS achieves universality through configuration, not through an additional layer of code abstraction.

This architectural decision renders the underlying C++ code universal and static. Consequently, GEARS can be adapted to an entirely new detector, source setup, or physics configuration without any modification or recompilation of the C++ source code. This makes GEARS a highly efficient application capable of rendering the complexity of many custom-written, medium-sized Geant4 applications unnecessary.

The universality of GEARS is facilitated by a set of rich, run-time configurable features, achieved through intelligent interfacing with the Geant4 macro command system:

\begin{description}
  \item[Detector Definition:] Utilizing plain text and GDML geometry I/O, allowing detector geometries and materials to be swapped out completely via simple files.
  \item[Physics Configuration:] Physics list selection is performed at launch via an environment variable, and individual process switching is configured through Geant4 macro commands.
  \item[Primary Source Definition:] Universal source configuration is driven entirely by Geant4 General Particle Source (GPS) macro commands.
  \item[Data Output:] Data members are named using generic conventions to be non-experiment-specific, and their recording can be dynamically turned on or off by adjusting detector component names or copy numbers in detector definition files.
\end{description}

The remainder of this paper details the unique computational architecture that enables this minimal code base and maximal functionality of GEARS. Section 2 discusses the design considerations that led to the minimalist approach. Section 3 outlines the distribution of the GEARS executable, and Section 4 provides acknowledgments.

\section{Application Versus Framework}
The minimal C++ footprint and simplified learning curve of GEARS stem directly from deliberate choices in its coding style, which prioritizes the conventions of a standalone executable application over those of a reusable software framework or library. While the official Geant4 examples often follow the extensive, defensive conventions of the core Geant4 toolkit itself (likely for instructional consistency), GEARS consciously deviates from this practice where it complicates readability and maintenance, demonstrating that a standalone application does not require a framework's structural overhead.

\subsection{Naming Conventions}
GEARS adopts concise, meaningful class names and avoids the use of custom capital letters or prefixes often used for identifier segregation in large libraries. For instance, the class handling the detector geometry is simply named \verb|Detector|, rather than adopting a convention such as \verb|GearsDetector|. This approach results in a highly readable and shorter codebase, and is justified because GEARS is designed as a single, standalone application, eliminating the need for complex, prefix-based identifier management.

\subsection{Single-File Architecture}
Traditional large software projects employ extensive directory structures and separate header (.hh or .h) and source (.cc or .cpp) files for organizational purposes. While effective for massive frameworks, this structure introduces overhead for applications, complicating code navigation and increasing the complexity of compilation systems (e.g., CMake). GEARS rejects this convention, consolidating all C++ code into a single source file. This choice fundamentally simplifies the build process and allows users to find all relevant definitions and logic in one place, greatly enhancing the ease of use for beginners and improving the speed of modification for experts.

\subsection{Namespace Strategy}
Furthermore, GEARS deliberately avoids the use of custom C++ namespaces. Namespaces are crucial for large software frameworks or libraries where multiple components may be loaded simultaneously, requiring a mechanism to prevent identifier collisions. Since GEARS is intended to be compiled and run as a single executable, it eliminates the need for namespace declarations. This choice reduces code volume and removes the cognitive load associated with qualifying identifiers (e.g., \verb|Gears::Detector::Construct()| versus the simpler \verb|Detector::Construct()|), making the code base significantly cleaner and easier for a beginner to follow and debug.

\subsection{Multiple Inheritance}
Another contraversal choice made for code minimization is the strategic use of multiple inheritance within key application classes. For example, the \verb|Detector| class is defined as:
\begin{lstlisting}[language=c++]
class Detector :
  public G4VUserDetectorConstruction,
  public G4UImessenger
\end{lstlisting}
In the context of a framework, inheriting from conceptually distinct base classes like \verb|G4VUserDetectorConstruction|, which constructs the detector in memory, and \verb|G4UImessenger|, which defines new macro commands, would typically be avoided to maintain a clean separation of concerns.
However, in the context of an application like GEARS, this allows for convenient coupling of the detector definition with new macro commands importing/exporting detector definition files, reducing the number of C++ classes and overall code required to link the two functionalities.

\section{Compilation and Installation}
GEARS utilizes the CMake build system to ensure native compilation and execution across the three major operating systems: Windows, macOS, and Linux. The CMakeLists.txt file is designed to reflect the single-file C++ architecture, which ensures that the build process remains fast and reliable across different environments.

A common hurdle for new users of complex physics software is locating the compiled executable and correctly setting up the necessary environment variables required by the toolkit (e.g., library paths, data files). GEARS addresses this through automated build and runtime steps. During the compilation process, the instructions within CMakeLists.txt automatically relocate the compiled executable (gears on Unix-like systems, or gears.exe on Windows) from the standard build directory (e.g., /path/to/gears/build/) to the application's root directory (/path/to/gears/). This prevents new users from encountering confusion when attempting to find the executable in a temporary build folder. For immediate and easy execution, the GEARS distribution provides shell scripts, gears.sh for Linux and macOS, and gears.bat for Windows. Crucially, on Windows, the gears.bat file is automatically called by the CMake build process to ensure the necessary environment variables are set immediately, acknowledging that a regular Windows user may not know how to manually execute a batch file. On all platforms, these scripts ultimately modify the user's local PATH environment variable to include the /path/to/gears/ directory, allowing users to execute the GEARS application from any location in their terminal or command prompt by simply typing gears. This deployment structure ensures that the overhead of installation and initial execution is minimal, further lowering the barrier to entry for beginners and increasing overall productivity.

\section{Detector Definition without C++}
\label{s:det}
The universal nature of GEARS, which allows it to simulate distinct experiments without C++ code modification, is rooted in the strategic implementation of configuration interfaces that maximize the run-time control over core Geant4 modules. Specifically, the necessity of hard-coding geometry in the derived \verb|G4VUserDetectorConstruction| C++ class is eliminated.

To achieve this, the \verb|Detector| class in GEARS utilizes its inheritance from \verb|G4UImessenger| to expose new Geant4 macro commands for detector definition file input/output (I/O), as shown in Listing~\ref{l:detio}. These commands allow users to define and place detector components using external text files in two distinct formats:
\begin{enumerate}
    \item Geometry Description Markup Language (GDML)~\cite{gdml06} files, identifiable by the suffix \verb|.gdml|.
    \item Plain Text Geometry (TG) files, using a simple text description syntax~\cite{tg} and the suffix \verb|.tg|.
\end{enumerate}
GEARS automatically determines the appropriate interface, either GDML or text geometry parser, to load the configuration based on the file's suffix.

\begin{lstlisting}[caption={New Geant4 macro commands for detector definition file I/O.}, label=l:detio, numbers=left, numbersep=5px, numberstyle=\color{blue}]
/geometry/source detector.tg
/run/initialize
/geometry/export detector.gdml
\end{lstlisting}

\subsection{GDML}
GEARS fully utilizes the standard Geant4 GDML parser. This method allows users to load complex detector geometries, materials, and associated physical properties from industry-standard XML files. GDML is conceptually similar to HTML. It is highly structured, verbose, and designed for machine-to-machine exchange, such as transferring geometries from CAD systems into Geant4, or from Geant4 to data analysis software, such as ROOT~\cite{root}. Its support in GEARS ensures broad compatibility with existing simulation workflows.

\subsection{Plain Text Geometry}
For fast prototyping and simple geometries, GEARS fully implements and showcases the use of a standard Geant4 functionality -- importing detector definition in plain text description. This capability, while documented in the Geant4 User's Guide~\cite{g4doc}, is often overlooked in favor of C++ class construction. In contrast to GDML, text geometry is conceptually similar to Markdown~\cite{md}. It is minimalist, easy to read, and optimized for human creation and modification. For example, an experimental hall filled with air and of a dimension of $10 \times 10 \times 10$ meters can be easily implemented using the following line:

\begin{lstlisting}
:volu hall BOX 5*m 5*m 5*m G4_AIR
\end{lstlisting}

For users constructing their geometry from scratch, this is the recommended approach due to its inherent simplicity and low cognitive load. By leveraging this existing Geant4 component, GEARS significantly lowers the barrier to entry, as users can define their geometry using an easy-to-read format without needing to master the complex C++ syntax or GDML.

It is worth noting that, GEARS includes a dedicated, non-standard extension to the built-in text geometry syntax for defining optical properties (e.g., reflectivity, absorption length, surface finishes). This extension is tightly integrated into the geometry loading mechanism and requires approximately 300 lines of dedicated C++ code within the single GEARS source file. This allows users to define and configure complex optical detectors (like scintillators or Cherenkov detectors) solely through external text files. The full details of this advanced functionality, including validation against experimental benchmarks, are provided in a separate, dedicated publication.

\subsection{Performance of Text Geometry I/O}
Using a text-based geometry file presents no performance penalty during the simulation run itself when compared to geometry defined directly in C++. This is because, during the Geant4 initialization phase, the text geometry parser translates the content of the external configuration file into the same underlying C++ objects used by manually coded definitions. While there is a minimal delay associated with reading and parsing the external text file during the application's startup, this effect is negligible and virtually never noticed in the daily use of GEARS. Therefore, the choice of text geometry offers significant configurability benefits without sacrificing simulation speed.

\subsection{Handling Complex Geometries}
While the plain text geometry syntax is simple, it can be seamlessly used to define highly complex detectors. When geometry becomes exceedingly complex, the challenge shifts from defining the syntax to managing the large volume of component placement and parameter data. To handle this, users can leverage any high-level programming language that excels at string manipulation, such as Python or shell scripts, to programmatically generate the text geometry definition files. A short loop within one of these scripts can generate a text file containing thousands of lines of detector placements. The simple, non-hierarchical syntax of the text geometry is particularly suited for this kind of programmatic generation, allowing users to create sophisticated detector models without ever needing to touch or compile C++ code. This enables external scripting to handle the complexity, reinforcing GEARS's core mission.

\subsection{Geometry Conversion}
The first line in Listing~\ref{l:detio} is used to load the geometry from an external file, the last line is used to write the currently loaded detector geometry from memory to a specified output file format. The intermediate command, \verb|/run/initialize|, is necessary to trigger the actual construction of the geometry in memory after loading from the source file.

By combining these two commands within a macro file, GEARS can be used as a utility for converting detector definitions between the simple text geometry format and the standardized GDML format. This inter-operability feature is vital for passing detector definitions from Geant4 to ROOT or other data analysis software.

\section{Physics Configuration without C++}
In conventional Geant4 applications, customizing the physics list, which defines the particles, processes, and cross-section models, often requires modifying and recompiling the dedicated C++ class derived from \verb|G4VUserPhysicsList|. GEARS ensures that users can specify and fine-tune the list without writing or recompiling any C++ code by leveraging existing Geant4 mechanisms detailed below.

\subsection{Physics List Selection}
For choosing a reference physics list, such as FTFP\_BERT or QGSP\_BERT, etc., GEARS exclusively uses the PHYSLIST environment variable, a mechanism provided and supported by the Geant4 toolkit itself. Before the executable, the user sets this variable to the name of the desired reference physics list:
\begin{lstlisting}
PHYSLIST=QGSP_BERT gears run.mac
\end{lstlisting}
Based on this setting, Geant4 dynamically instantiates the corresponding validated physics models. Without this setting, GEARS will use the default physics list, \verb|FTFP_BERT|.

\subsection{Process Fine-Tuning}
Users can also turn individual physics processes or model sets on or off using existing Geant4 macro commands. Examples include \verb|/physics_list/factory/addXXX|, where XXX specifies a physics module like Optical or RadioactiveDecay, etc., and \verb|/process/activate| or \verb|/process/inactivate|, etc.

\subsection{Advanced Radioactive Decay Management}
A key challenge in standard Geant4 radioactive decay simulation is that the entire decay chain, regardless of the real-world time elapsed between decays, is simulated within a single event. This makes isolating or analyzing daughter nuclear decays that occur significantly later in real time difficult, as they are not properly separated into distinct events.

GEARS addresses this by introducing a dedicated macro command for event separation. Approximately 40 lines of C++ code in gears.cc are used to define this new command, which allows the user to specify a time window (e.g., 1 second):
\begin{lstlisting}
/grdm/setTimeWindow 1 s
\end{lstlisting}
Geant4 simulation provides the time information for when a decay occurs. If the decay time of a daughter nucleus is longer than the set time window, the application splits the decay chain: the current event is terminated, and the rest of the decay chain is saved to be simulated as a new, separate Geant4 event. This feature is crucial for efficient event separation and streamlined analysis in applications involving delayed coincidences or complex isotope production.

\section{Primary Particle Source Definition without C++}
To achieve complete run-time configurability for the primary particle source, from which a Geant4 simulation starts, GEARS accepts two sets of standard Geant4 source macro commands: General Particle Source (GPS) commands (/gps/...) and the simple gun commands (/gun/...). 
The formers are more flexible and enabling the definition of complex source parameters, including advanced energy spectra, angular distributions, and spatial geometries (e.g., cylindrical, planar). The later are simpler and allows users to easily configure basic sources, such as mono-energetic point sources, providing a low-overhead option for simple simulation setups and making the application accessible to beginner Geant4 users. This architectural choice ensures that the source configuration remains entirely decoupled from the application's static C++ core, reinforcing the universality of GEARS.

\section{Data Management and Output}
Generally speaking, the visualization of detector geometry and particle trajectories, and the screen dump of a Geant4 application triggered by \verb|/tracking/verbose 2| can all be regarded as output of a Geant4 simulation. Strictly speaking, the output of a Geant4 simulation includes histograms and/or ntuples (multi-dimensional spreadsheets) generated during the simulation, which can be used to reveal statistical distributions of, for example, positions and energy depositions of interactions. GEARS introduces several design ideas to its data output system for optimized ntuple analysis.

\subsection{Data Format}
The choice of output file name and format in GEARS is fully configured at run-time using the macro command:
\begin{lstlisting}
/analysis/setFileName output.root
\end{lstlisting}

One of the following suffixes is used to specify the desired output file format: .root, .hdf5, .csv, or .xml. Note that the output is disabled by default. It will be enabled if the output file name is not empty, meaning this macro command also serves as the output switch. Without it, no output file will be created.

Due to the following computational and analytical advantages, the ROOT format (\verb|.root|) is recommended for GEARS output data management.
\begin{description}
    \item[Scalability \& Memory Efficiency:] The format is specifically designed to work with large data sets. It achieves this by loading individual data branches (similar to columns in a spreadsheet) separately to computer memory for analysis, which requires a relatively small memory footprint compared to loading the entire data set at once, as many other analysis tools do.
    \item[Compression \& Storage:] The format compresses data to save disk space, resulting in a smaller file size compared to pure text formats, such as CSV or XML.
    \item[Built-in Analysis Tools:] It provides native functions to create and analyze statistical distributions of data, including multiple-dimensional histograms, simplifying the initial analysis steps. Furthermore, it leverages the power and simplicity of the \verb|TTree::Draw()| functionality for rapid data querying.
    \item[Integration \& Interoperability:] the ROOT output format is a built-in feature of Geant4, meaning no external libraries (like HDF5) are needed for core functionality, simplifying the installation and deployment process.
    \item[Cross-Language Interoperability:] The data remains accessible to modern data analysis ecosystems, including Python and Julia, through widely supported packages like uproot.
\end{description}

\subsection{Step Point Information}
GEARS focuses on saving detailed information for every discrete computational step of a particle's trajectory, which are termed step points. A comprehensive list of variables is saved per step point, utilizing short (maximally three letters) names to simplify analysis queries as shown in Table \ref{t:spv}. While these abbreviated names (e.g., \verb|trk| for Track ID) are initially harder to interpret compared to verbose long names, the resulting gain in interactive analysis speed and typing efficiency is substantial. This drawback is overcome through detailed documentation, such as Table \ref{t:spv}, which provides the precise mapping between the short variable name and its full physical description.
\begin{table}[t]\centering
\caption{List of step point variables.} \label{t:spv}
\begin{tabular}{llc}
\toprule
\textbf{Short} & \textbf{Full} & \textbf{Variable} \\
\textbf{Name} & \textbf{Description} & \textbf{Unit} \\
\midrule
trk & Track ID & N/A \\
stp & Step number (starting from 0) & N/A \\
vlm & Detector volume copy number & N/A \\
pro & Process ID (see below) & N/A \\
pdg & Particle ID (PDG encoding) & N/A \\
pid & Track ID of the parent particle & N/A \\
xx & Local position X & [mm] \\
   & (origin: center of the volume) &  \\
yy & Local position Y & [mm] \\
   & (origin: center of the volume) & \\
zz & Local position Z & [mm] \\
   & (origin: center of the volume) & \\
dt & Local time (time elapsed from & [ns] \\
   & previous step point & \\
de & Energy deposited & [keV] \\
dl & Step length & [mm] \\
l & Trajectory length & [mm] \\
x & Global position X & [mm] \\
  & (origin: center of the world) & \\
y & Global position Y & [mm] \\
  & (origin: center of the world) & \\
z & Global position Z & [mm] \\
  & (origin: center of the world) & \\
t & Global time (time elapsed from & [ns] \\
   & the beginning of event) & \\
k & Kinetic energy of the particle & [keV] \\
p & Momentum of the particle & [keV] \\
q & Charge of the particle & N/A \\
\bottomrule
\end{tabular}
\end{table}

\subsection{Process ID Encoding}

To facilitate simple plotting and event selection, GEARS consistently saves numerical identifiers instead of verbose string names. For instance, it saves a detector volume's copy number (\verb|vlm|) instead of its name, and a particle's PDG encoding (\verb|pdg|) instead of its name. Following this same idea, GEARS converts the verbose Geant4 physics process names (e.g., ``Transportation'', ``Photoelectric'', etc.) into an integer (Process ID). The physics process generating each step point is saved in the variable \verb|pro[i]|, where \verb|i| is the index of the step point.

The Process ID is encoded using the following formula, containing both the general process category and a specific sub-type:
$$
\text{Process ID} = (\text{Process Type}) \times 1000 + (\text{Sub-Type}).
$$
Major process types are defined in file \verb|G4ProcessType.hh| in Geant4 source code, and sub-types are defined in 
\begin{itemize}
    \item \verb|G4HadronicProcessType.hh|
    \item \verb|G4DecayProcessType.hh|
    \item \verb|G4EmProcessSubType.hh|
    \item \verb|G4TransportationProcessType.hh|
    \item \verb|G4FastSimulationProcessType.hh|
    \item \verb|G4OpProcessSubType.hh|
\end{itemize}
The corresponding process IDs are listed below:

\begin{itemize}\small
    \item less than 1000: not defined
    \item 1000 to 2000: Transportation
    \begin{itemize}
        \item 1000: initial step (step 0)
        \item 1091: transportation
        \item 1092: coupled transportation
    \end{itemize}
    \item 2000 to 3000: Electromagnetic
    \begin{itemize}
        \item 2001: Coulomb scattering
        \item 2002: ionization
        \item 2003: Bremsstrahlung
        \item 2004: pair production by charged particles
        \item 2005: annihilation
        \item 2010: multiple scattering
        \item 2011: Rayleigh scattering
        \item 2012: photoelectric effect
        \item 2013: Compton scattering
        \item 2014: gamma conversion (pair production)
        \item 2016: gamma general process (2010 to 2014)
        \item 2021: Cherenkov
        \item 2022: scintillation
        \item 2023: synchrotron radiation
    \end{itemize}
    \item 3000 to 4000: Optical
    \begin{itemize}
        \item 3031: absorption
        \item 3032: boundary
        \item 3033: Rayleigh scattering
        \item 3034: WLS (Wavelength Shifting)
        \item 3035: Mie scattering
        \item 3036: WLS2 (Wavelength Shifting 2)
    \end{itemize}
    \item 4000 to 5000: Hadronic
    \begin{itemize}
        \item 4111: hadron elastic
        \item 4121: hadron inelastic
        \item 4131: capture
        \item 4132: muon atomic capture
        \item 4141: fission
        \item 4151: hadron decay at rest
        \item 4142: lepton decay at rest
        \item 4161: charge exchange
        \item 4210: radioactive decay
    \end{itemize}
    \item 5000 to 6000: Photolepton\_Hadron
    \item 6000 to 7000: Decay
    \begin{itemize}
        \item 6201: decay
        \item 6202: decay with spin
        \item 6203: pion decay with spin
        \item 6210: radioactive decay
    \end{itemize}
    \item 7000 to 8000: General
    \begin{itemize}
        \item 7403: neutron killer
    \end{itemize}
    \item 8000 to 9000: Parameterisation
    \item 9000 to 10000: User Defined
    \item 10000 to 11000: Parallel
    \item 11000 to 12000: Phonon
    \item 12000 to 13000: UCN (Ultra-Cold Neutron)
\end{itemize}

This structured encoding ensures that users can select events generated by a specific process (e.g., photoelectric effect) using simple integer comparisons rather than complex string matching in the analysis tools.

\subsection{Flat VS Nested Ntuples}
Conventional Geant4 applications often organize their output data in a deeply nested ntuples, reflecting the object-oriented hierarchy of events, tracks, steps, and various physics quantities associated with each step. While architecturally clean, this results in verbose variable paths that severely hinder interactive analysis.

GEARS implements a radical departure by producing a flat ntuple using the ROOT \verb|TTree| class, which is commonly called a tree. Such a flat tree structure can be thought of as a spreadsheet. Each row is called an entry (or an event) and each column is called a branch (or a leaf). If you simulate 1000 events using GEARS, your ROOT tree would have 1000 entries. If you save the first step point position $x$, $y$, and $z$ in the tree, your tree would have 3 branches, each holding 1000 values of $x$, $y$, or $z$. If an event has more than one step point, you will have a few $x$, $y$, and $z$'s for each event (row), and your table would have one more dimension, indexed by variable \verb|Iteration$| in \verb|TTree|. The depth of this dimension may change event by event since each event may have a different number of step points. It is saved in a branch called \verb|n|, which is the number of step points in each event. The tree itself is saved in the output file as a single letter pointer \verb|t|. This flat structure and short variable names dramatically simplify analysis query language.

Compare the complexity of accessing a variable in a typical nested structure
\begin{lstlisting}[language=c++, otherkeywords={Draw}]
t->Draw("event.track(1).step(3).momentum[1]")
\end{lstlisting}
versus a flat structure using the \verb|TTree::Draw()| method:
\begin{lstlisting}[language=c++, otherkeywords={Draw}]
t->Draw("py", "trk==1 && stp==3")
\end{lstlisting}
where the variable in the first pair of quotes is the one to be plotted in a histogram and expressions in the second pair of quotes are selection conditions.

The efficiency of the flat structure is most pronounced when executing complex analyses. For instance, plotting the stopping power of the primary particle within a detector volume using the flat structure requires only the following short command:
\begin{lstlisting}[language=c++, otherkeywords={Draw}]
t->Draw("de/dl", "trk==1")
\end{lstlisting}
In contrast, achieving the same result using a standard nested structure would necessitate an unbearably long command:
\begin{lstlisting}[language=c++, otherkeywords={Draw}]
t->Draw("event.track(1).step().deltaE/"
        "event.track(1).step().deltaL")
\end{lstlisting}

\subsection{Step Zero}
The G4VUserSteppingAction abstract class is the recommended interface for users to get access to information associated with individual steps of a particle's propagation through the implemented detector geometry. Interestingly, step zero, or the initial step (initStep printed on screen if /tracking/verbose is set to be more than 0), is not accessible through this interface.

However, step zero contains some very important information that is constantly requested by the user. For example, the energy of a gamma ray from a radioactive decay is only available from step zero. Such information can be easily displayed using the following ROOT command with the Output ROOT tree, t:
\begin{lstlisting}[language=c++, otherkeywords={Draw}]
t->Draw("k","pro==6210 && pdg==22")
\end{lstlisting}
which draws the kinetic energy \verb|k| of a $\gamma$ (\verb|pdg|==22) created by the radioactive decay process (\verb|pro|==6210).

This is achieved by using \verb|G4SteppingVerbose| instead of \verb|G4VUserSteppingAction| for data recording. The former provides a function \verb|TrackingStarted()| to print information of step zero on screen if /tracking/verbose is set to be more than 0. It also provides a function \verb|StepInfo()| to print information about steps after step zero on screen if /tracking/verbose is more than 0. Since \verb|StepInfo()| and \verb|UserSteppingAction()| are both called in \verb|G4SteppingManager::Stepping()|, the former can be used to fully replace the functionality of the latter. Specifically, the \verb|Output| class in GEARS inherits class \verb|G4SteppingVerbose|, and overwrites its member functions \verb|TrackingStarted()| and \verb|StepInfo()| to record data from step zero and the rest steps, respectively.

There is another advantage of using \verb|G4SteppingVerbose| instead of \verb|G4UserSteppingAction| for data recording, that is, \verb|G4SteppingVerbose| is provided as a globally available singleton, which can be easily accessed at different places in the codes using:
\begin{lstlisting}[language=c++]
G4VSteppingVerbose::GetInstance()
\end{lstlisting}
It is used in \verb|G4UserRunAction| to open and close the output file, in \verb|G4UserEventAction| to fill the output tree.

Note that functions in \verb|G4SteppingVerbose| will not be called in \verb|G4SteppingManager| unless /tracking/verbose is set, which will print too much information on screen for a long run. This is solved in \verb|EventAction::BeginOfEventAction| by turning on tracking verbose all the time so that all functions in \verb|G4SteppingVerbose| will be called, while at the same time, turning on \verb|G4SteppingVerbose| local verbose flag \verb|Silent| to run them in silent mode.

\subsection{Dynamic Data Logging Control}
GEARS uses simple strings to control data logging on the fly.
A user can designate any volume (e.g., a volume named ``detector'') as sensitive simply by appending the string ``(S)'' to the end of its name within the configuration file, resulting in ``detector(S)''.

For any volume designated as sensitive, the total energy deposited within that volume is automatically summed up and saved into an array variable, \verb|et[m]|, where \verb|m| is the copy number of the volume. The first variable in the array \verb|et[0]| is reserved to store the sum of all energy deposits across all other sensitive volumes.

It is important to note that currently, only the total energy deposit is calculated for a sensitive volume. However, the modular and non-intrusive nature of this naming system ensures that other quantities (e.g., particle counts, doses, etc.) related to a sensitive volume can be added and recorded easily in future versions of GEARS if needed.

For completeness, the functionality of energy deposition tallying in specific volumes can also be achieved using the official Geant4 scoring commands, such as those under the \verb|/score/| macro command set, which provides tallies for other quantities as well, such as particle counts, radiation doses, etc. GEARS fully supports and encourages the use of these official commands as they perfectly align with the core GEARS philosophy: enabling users to perform advanced output configuration entirely at run-time through macro commands, without requiring any modifications or recompilation of the C++ source code. The GEARS-specific \verb|(S)| string mechanism is provided as a simple alternative for one of the most commonly requested quantity: total energy deposition.
    
The recording of granular step point data for a specific detector volume can be dynamically controlled based on its copy number. Assigning a copy number greater than 0 turns step point data recording in that volume ON, while assigning a copy number less than or equal to 0 turns recording OFF. Note that the copy number of the top most volume (typically called the ``world'') is 0. No data is then recorded for the world volume. This dual control allows users to precisely tailor the output file size and content to their analysis needs.

\subsection{Analysis Snippets}

The combination of the flat ntuple data structure and the concise variable names is designed to maximize interactive analysis efficiency. The power of this design is best demonstrated through concrete examples of data querying using the ROOT \verb|TTree::Draw()| command, which executes analysis logic directly on the data file without requiring a compiled analysis script. The following examples illustrate common analysis tasks performed directly within a ROOT session, where \verb|t| is the pointer to the GEARS output \verb|TTree|.

\begin{lstlisting}[language=c++]
t->Draw("x:y", "trk==1", "l", 1, 0)
\end{lstlisting}
draws tracks of the primary particle (\verb|trk==1|) on the $x$-$y$ plane. The \verb|"l"| option forces the output to be drawn as lines, and the final arguments, \verb|1, 0|, instruct ROOT to process only the first entry (event) in the tree. It demonstrates that the GEARS output structure can be used not only for statistical analysis, but also for individual event visualization.

\begin{lstlisting}[language=c++]
t->Draw("pro", "trk>1 && stp==0")
\end{lstlisting}
shows which physics process (\verb|pro|) are involved in generating (\verb|stp==0|) secondary particles (\verb|trk>1|). 

\begin{lstlisting}[language=c++]
t->Draw("x:y:z", "vlm==1")
\end{lstlisting}
displays the three-dimensional spatial distribution of all step points ("hits") within the detector volume whose copy number (\verb|vlm|) is 1.

\begin{lstlisting}[language=c++]
t->Draw("pdg")
\end{lstlisting}
provides a quick overview of the types of particles that were tracked.

\begin{lstlisting}[language=c++]
t->Draw("pro", "pdg==22 && stp!=0")
\end{lstlisting}
plots physics processes related to $\gamma$-rays (\verb|pdg==22|), but not the generation of $\gamma$-rays (\verb|stp!=0|).

\begin{lstlisting}[language=c++]
t->Draw("x:y", "trk>1")
\end{lstlisting}
plots the $x$-$y$ spatial distributions of secondary particles (\verb|trk>1|) throughout the simulated geometry.

\begin{lstlisting}[language=c++]
t->Draw("de/dl:p")
\end{lstlisting}
plots the energy loss per unit length (\verb|de/dl|), which represents the stopping power, versus the particle's momentum (\verb|p|). This is a fundamental plot for validating energy deposition models.

\begin{lstlisting}[language=c++]
t->Draw("et[1]")
\end{lstlisting}
draws the total energy spectrum recorded by the sensitive volume (detector) with copy number 1, stored in the dedicated energy tally variable \verb|et[1]|.

\section{Documentation and Dissemination}

A high-quality software package requires accessible documentation and multiple avenues for user engagement. GEARS utilizes a multi-platform strategy to ensure comprehensive technical documentation, tutorial support, and user outreach, leveraging modern collaborative tools and multimedia platforms.

\subsection{Code Hosting}

The codebase for GEARS is hosted on GitHub~\cite{github}, a platform that inherently supports streamlined documentation. This feature is fully utilized by placing descriptive \verb|README.md| files in all major directories and subdirectories of the repository. GitHub automatically renders these Markdown files into nicely formatted webpages, providing immediate, context-specific documentation for users browsing the file structure. This approach ensures that technical instructions and usage examples are intrinsically linked to the relevant code components.

\subsection{Automated Documentation}

For developers seeking deep insight into the internal C++ variables, methods, classes, and their relationships, the GEARS repository is connected to CodeDocs.xyz~\cite{codedocs}. which automatically processes the source code in \verb|gears.cc| using Doxygen~\cite{doxygen}, generating comprehensive, up-to-date documentation.

\subsection{Multimedia User Tutorials}

Recognizing that visual and audio content often simplifies complex technical concepts, the GEARS project actively disseminates usage tutorials through the Physino YouTube channel~\cite{physino}. Videos on this platform demonstrate the advantages and practical usage of the software, making complex Geant4 concepts easier to understand than pure text manuals. By offering this multimedia approach, the Physino channel has established itself as one of the most popular YouTube resources dedicated to Geant4 simulation, significantly lowering the barrier to entry for new GEARS users.

\subsection{User Support and Community Engagement}

In addition to static documentation, the GEARS project maintains active channels for direct user support and feedback. Users are encouraged to ask questions and report issues through two primary platforms:
\begin{description}
    \item[GitHub Issue Tracker.] The Issues tab on the GitHub repository is the official channel for reporting bugs, suggesting features, and submitting technical queries.
    \item[YouTube Comments.] The comments section of the Physino YouTube channel serves as a dynamic, informal forum where users can ask questions about specific tutorials or general usage, fostering community interaction and rapid response to common queries.
\end{description}
These open channels ensure that the project remains responsive to the community and provides iterative support based on real-world user needs.

\section{Executable Distribution and Usability}

The GEARS project tackles a significant barrier to entry for new Geant4 users: the complexity of compilation of not only Geant4 libraries but also the GEARS executable. It addresses this challenge through a deliberate choice of single-threaded architecture and modern executable distribution via containerization.

\subsection{Addressing the Executable Barrier via Docker}

Geant4 is primarily distributed as a framework consisting of a set of development toolkits and libraries, rather than a final, double-clickable executable product. This poses a major adoption barrier, especially for beginners and users on platforms like Windows, who expect a ready-to-launch application.

To provide a true, instantly usable product, GEARS is distributed within a Docker image hosted on Docker Hub~\cite{g4img}. This container image contains the official Geant4 libraries, pre-compiled on AlmaLinux, along with the GEARS executable compiled against them.

This approach provides a ready-to-use, standardized platform that offers several advantages:

\begin{itemize}
    \item Universal Compatibility: It provides a consistent execution environment for users on all major operating systems (Windows, macOS, Linux).
    \item Ease of Deployment: It allows end-users to launch the GEARS executable immediately after a simple Docker pull command, completely bypassing the complex process of compiling Geant4 libraries and application code.
    \item Cloud Readiness: The containerized nature inherently enables simple deployment and scaling in future cloud computing environments.
\end{itemize}

\subsection{Single-Threaded Design for Beginner Accessibility}

GEARS is intentionally provided as a single-threaded application. While Geant4 supports multi-threading for computational speed, the resulting output structure can be complex and challenging for a new user to interpret, often leading to data that is perceived as disorganized when trying to follow individual events or steps.

By adhering to a single-threaded design, GEARS prioritizes beginner-friendliness and output clarity, ensuring that the tracking and event information is sequential and easy to understand.

\subsection{Utilizing Multi-Instance Parallelization}

The choice of single-threaded execution does not sacrifice the ability to utilize powerful modern CPUs. Instead of relying on internal multi-threading, GEARS leverages the power of containerization for horizontal scaling. Full use of a powerful CPU is achieved by running multiple executables or container instances of the single-threaded GEARS application in parallel. This approach is cleaner and more robust for data separation, as each instance generates its own clean, single-threaded output file. The process is managed externally by the operating system or orchestration tools, avoiding the internal data management complexity associated with Geant4 multi-threading for the beginner user.

\section{Conclusion}

The primary objective of the GEARS project is to transform the Geant4 toolkit from a framework requiring C++ development into a fully self-contained, run-time configurable simulation application. The comprehensive design choices detailed throughout this paper successfully meet this goal, making GEARS a highly accessible and efficient platform for particle transport simulation across a vast array of experimental applications.

This efficiency is rooted in four key pillars of design, all centered on eliminating the need for C++ modification:

\begin{itemize}
    \item Geometry Configurability: The adoption of text-based geometry definition makes design and modification entirely external to the C++ core, supporting complex detector configurations with immediate portability.
    \item Physics Configurability: The strict use of the Geant4 factory method via the \verb|PHYSLIST| environment variable ensures validated physics selection through simple external commands.
    \item Primary Particle Configurability: Exclusive reliance on the General Particle Source (GPS) and traditional \verb|/gun| macros ensures the primary particle characteristics are dynamically configured via text files.
    \item Output Configurability: Output control is managed entirely at run-time. The creation of the output file is controlled by a macro command. The granularity of the recorded data is controlled per detector volume using the volume's copy number, while energy tallying for a volume is enabled simply by attaching the \verb|(S)| string to the volume's name.
\end{itemize}

Beyond run-time control, the approach of treating GEARS as a complete application instead of a framework allows for several key optimizations that enhance the user experience. This includes the flat Ntuple data structure, coupled with short, intuitive variable names, which dramatically streamlines the post-simulation analysis workflow, and the use of \verb|G4SteppingVerbose| to capture necessary step zero data provides a complete record of particle interactions.

In summary, GEARS provides a ready-to-use, powerful simulation platform where the entire workflow, from setting up a new detector and selecting the physics models to defining the particle source and performing complex data analysis, can be performed for most small to medium-size experiments without requiring any modification or recompilation of C++ code. This design lowers the barrier to entry for Geant4 and significantly accelerates the prototyping and analysis cycles for the scientific community.

\section{Acknowledgements}
This work is supported by the NSF award OIA-2437416, PHY-2411825, and the Office of Research at the University of South Dakota. Computations supporting this project were performed on High Performance Computing systems at the University of South Dakota, funded by NSF award OAC-1626516.

\section{Declaration of generative AI and AI-assisted technologies in the manuscript preparation process}

During the preparation of this work the author used \texttt{Gemini} in order to generate the initial draft of this manuscript. After using this tool, the author reviewed and edited the content as needed and takes full responsibility for the content of the published article.

\bibliographystyle{elsarticle-num}
\bibliography{ref}
\end{document}